\documentclass{aastex}
\usepackage{emulateapj5,apjfonts,psfig,epsfig}

\def\asca{{\sl ASCA }}
\def\bepposax{{\sl BeppoSAX }}

\def\ergsec{\hbox{erg s$^{-1}$ }}
\def\ergcm{\hbox{erg cm$^{-2}$ s$^{-1}$ }}

\slugcomment{Accepted for publication in The Astrophysical Journal}

\shorttitle{X-RAY LINES IN 4U 1626$-$67}

\shortauthors{SCHULZ ET. AL.}

\begin{document}

\title{DOUBLE-PEAKED X-RAY LINES FROM THE OXYGEN/NEON-RICH ACCRETION
DISK IN 4U 1626$-$67} 
\author{
Norbert~S.~Schulz,
Deepto~Chakrabarty,\altaffilmark{1}
Herman~L.~Marshall,
Claude~R.~Canizares,\altaffilmark{1}
Julia~C.~Lee,
and
John~Houck
 }
\affil{Center for Space Research, Massachusetts Institute of Technology,
Cambridge, MA 02139; nss,deepto,hermanm,crc,jlee,houck@space.mit.edu}

\altaffiltext{1}{Also Department of Physics, Massachusetts Institute of
Technology, Cambridge, MA 02139.}

\begin{abstract}
We report on a 39 ks observation of the 7.7-s low-mass X-ray binary
pulsar 4U~1626$-$67 with the High Energy Transmission Grating
Spectrometer (HETGS) on the {\em Chandra X-Ray Observatory}.  This
ultracompact system consists of a disk-accreting magnetic neutron star
and a very low mass, hydrogen-depleted companion in a 42-min binary.
We have resolved the previously reported Ne/O emission line complex
near 1~keV into Doppler pairs of broadened ($\approx$2500 km s$^{-1}$
FWHM) lines from highly ionized Ne and O.  In most cases, the
blue and red line components are of comparable strength, with
blueshifts of 1550--2610 km~s$^{-1}$ and redshifts of 770--1900
km~s$^{-1}$.  The lines appear to originate in hot ($\approx 10^6$ K),
dense material just below the X-ray--heated skin of the outer
Keplerian accretion disk, or else possibly in a disk wind driven from
the pulsar's magnetopause.  The observed photoelectric absorption
edges of Ne and O appear nearly an order of magnitude stronger than
expected from interstellar material and are likely formed in cool,
metal-rich material local to the source.  Based on the inferred local
abundance ratios, we argue that the mass donor in this binary is
probably the 0.02 $M_\odot$ chemically fractionated core of a C-O-Ne or
O-Ne-Mg white dwarf which has previously crystallized.  
\end{abstract}

\keywords{
accretion, accretion disks ---
binaries: close ---
stars: individual (4U 1626$-$67) ---
stars: neutron ---
techniques: spectroscopic ---
X-rays: stars}

\section{Introduction}

The ultracompact low-mass X-ray binary 4U 1626$-$67 consists of a
7.7~s pulsar accreting from an extremely low-mass ($\lesssim 0.1
M_\odot$) companion with a binary separation of only $\approx$1
lt-sec.  The very short binary period ($P_{\rm orb}$=42 min;
Middleditch et al. 1981; Chakrabarty 1998) indicates that the
companion must be extremely hydrogen-poor (Paczynski \& Sienkiewicz
1981; Nelson, Rappaport, \& Joss 1986).  Based on X-ray timing limits,
the Roche-lobe--filling companion must be either a 0.02 $M_\odot$ white
dwarf or a 0.08 $M_\odot$ hydrogen-depleted, partially degenerate star
(Levine et al. 1988; Verbunt, Wijers, \& Burm 1990; Chakrabarty 1998).
These solutions correspond to relatively face-on binary inclinations
of 33$^\circ$ and 8$^\circ$, respectively.  For a randomly oriented
ensemble of binaries, the a priori probabilities of finding such low
inclinations are only 16\% and 1\%, respectively\footnote{A more
massive 0.6 $M_\odot$ helium-burning star is also allowed by the X-ray
timing limits (with $i=1.3^\circ$ and an a priori probability of
0.03\%), but this solution requires an implausible (36 kpc) distance
to be consistent with the observed X-ray flux.}.

The pulsar's spin frequency evolves on short ($|\nu/\dot\nu|\approx
6000$ yr) time scales due to accretion torques.  After spinning up
steadily for 13 years, the accretion torque abruptly changed sign in
1990, leaving the pulsar in a prolonged spin-down state (Chakrabarty
et al. 1997).   The surface dipole magnetic field strength of $3\times
10^{12}$ G, inferred from the detection of cyclotron absorption
lines (Orlandini et al. 1998), suggests that the pulsar is near spin
equilibrium.  The ultraviolet spectrum of the source is consistent
with that of an accretion disk whose inner edge is truncated by the
pulsar's magnetosphere near the corotation radius $r_{\rm co}=(GM_{\rm
X}P^2_{\rm spin}/4\pi^2)^{1/3}=6.5\times 10^8$ cm, and whose outer
edge is truncated at the binary's tidal radius $r_{\rm t}\approx
2\times 10^{10}$ cm (Wang \& Chakrabarty 2001).

The pulsar's X-ray continuum spectrum is evidently correlated with its
torque state.  During the 1977--1990 spin-up epoch, the pulsar's
pulse-phase-averaged X-ray spectrum was roughly described by the sum
of three separate power-law components with photon indices\footnote{We
define photon index $\gamma$ such that the photon flux $dN/dE\propto
E^{-\gamma}$.} $\gamma_1\approx1.4$--1.65 (0.7--10 keV),
$\gamma_2\approx0.5$ (10--20 keV), and $\gamma_3\approx 5$ (20--60
keV) (Pravdo et al. 1979; Maurer et al. 1982; Elsner et al. 1983; Kii
et al. 1986; Mavromatakis 1994).  However, observations acquired after
the 1990 torque reversal measured a significantly flatter soft X-ray
spectrum with $\gamma_1\approx0.6$--0.8 in the 1--10 keV range
(Angelini et al. 1995; Owens et al. 1997; Vaughan \& Kitamoto 1997).   
Since the discovery of this X-ray source over two decades ago, its
overall X-ray flux has declined steadily (Chakrabarty et al. 1997 and
references therein; Owens et al. 1997), but the optical flux from the
accretion disk has remained essentially constant over the same time
interval (Chakrabarty 1998).   

X-ray spectroscopy with \asca and \bepposax also detected a strong,
unusual complex of emission lines near 1~keV; these were identified
as the Ly$\alpha$ lines of hydrogenic and He-like Ne and O (Angelini
et al. 1995; Owens et al. 1997), with possible contributions from
Fe-$L$ shell emission. Angelini et al. (1995) also detected a faint
feature near 1.41 keV, which suggests the presence of photoionized
plasma if the feature is interpreted as the recombination continuum of
hydrogenic Ne.  Independent of whether the line complex arises
from a photoionized or a collisionally ionized plasma, however, its
strength indicates an overabundance of both Ne and O relative to
solar values (Angelini et al. 1995).  

In this paper, we present high-resolution spectra of 4U 1626$-$67
obtained with the High Energy Transmission Grating Spectrometer
(HETGS; Canizares et al. 2001, in preparation) on board the {\em
Chandra X-ray Observatory.}  

\section{Chandra Observations and Data Reduction}

4U 1626$-$67 was observed with the HETGS on 2000 September 16 (05:47:21
UT) continuously for 39 ks.  The HETGS carries two different
types of transmission gratings: the Medium Energy Gratings (MEG) and
the High Energy Gratings (HEG).  These allow for high-resolution X-ray 
spectroscopy in the 1--35 \AA\ (0.35--12.4 keV) range, with a spectral
resolution of $\lambda/\Delta\lambda \approx 1400$ at 12 \AA\ (1 keV)
and $\lambda/\Delta\lambda \approx 180$ at 1.8 \AA\ (6.9 keV). The
dispersed spectra were recorded with an array of 6 charged coupled
devices (CCDs) in the focal plane which are part of the Advanced CCD Imaging
Spectrometer (ACIS; Garmire et al. 2001, in preparation).  We defer to
the {\em Chandra} X-ray Center (CXC) documents for a more detailed
description of the spectroscopic instruments (see {\em
http://chandra.harvard.edu}). 

We recorded a total of 62373 events in the 1st order MEG spectrum and
33744 events in the HEG spectrum after applying standard event grade
selection cuts.  For some of the analysis, we combined the MEG and 
HEG spectra where their passbands overlapped.  In these cases, we used a
count-weighted average of the resolution of both gratings, which is
between 0.01 and 0.02 \AA.  The CXC provided aspect-corrected ``level 1''
event lists via standard pipeline processing.  These data were 
reprocessed using the latest available data processing input products.
We also removed all events that were attributed to bad pixels, bad
columns, or flaring background events that were not handled by the CXC
standard processing. 

The determination of the zeroth order source position is crucial for the
calibration of the wavelength scales, since it defines the zero
point for wavelength in the dispersed spectra.  We can determine the
source position on the detector to an accuracy of 0.2 pixels, which
corresponds to a wavelength zero point accuracy of 0.002 \AA\ in MEG
and 0.001 \AA\ in the HEG 1st order.  The current status of the
overall wavelength calibration is of order $\sim$ 0.05$\%$, leading to
a worst-case uncertainty of 0.004 \AA ~in the 1st order MEG at
\ion{O}{7} (21.6 \AA), and 0.006 \AA ~in the 1st order HEG at
\ion{Ne}{10} (12.1 \AA).  We processed the event lists into ``level 1.5''
grating event lists using available CXC software.  For the spectral
analysis we used a combination of custom software, FTOOLS, and the CIAO
2.1 package from CXC.  We computed aspect-corrected exposure maps for
each spectrum, allowing us to correct for effects from the effective
area of the CCD spectrometer. The overall flux calibration throughout
the CCD array is of order 10$\%$, although it can be as poor as 20$\%$
at lower energies in some areas.

The zeroth order image was affected by heavy ``pileup'', where the
event rate was so high that two or more events were detected in the
CCD during a 3.2 s frame exposure.  The time between frames was
$dt=3.24104$ s, which accounts for the time to transfer charge to the frame
store.  Pileup distorts the CCD spectrum because detected events
overlap so their deposited charges are collected into single,
apparently more energetic events.  More significant, however, is that
many events are lost on-board as the grades of the piled events
overlap those of rejected background.  In this case $\sim$ 90\% of the
events are lost.  We ignore the zeroth order events in all subsequent
analyses.

MEG and HEG events were selected for timing analysis on the basis of
spatial, pulse height, and grade selection (as described in the last
section).  Events from first and second orders were included, giving
102,835 events.  For the same pulse height and grade selection, a
background region 12.6\arcsec\ away gave only 386 events in the same
spatial window, so background is ignored.  Event times were corrected
for the motion of the spacecraft and then to the solar system
barycenter.  A random quantity uniformly distributed between 0 and
$dt$ was added to event times in order to avoid aliasing with the
sample time.  A significant peak was found in the FFT at a period of
7.672 s and a $\chi^2$ period folding method gave a best period of
7.6717(7) s, consistent with an extrapolation of the source's
long-term spin-down behavior (Chakrabarty et al. 1997).  The pulse appears
sinusoidal with an amplitude of 4.0 $\pm$ 0.9\%, averaged over the
entire bandpass. This sinusoidal appearance may as well be a consequence
of the fact that the pulse period is very close to the time resolution
of the detector. It has been shown in previous observations that
the shape and amplitude of the intrinsic pulse chnage with energy
(Pravdo et al. 1979, Kii et al. 1986, Owens et al. 1997). In this respect
we observe that the modulation of the events with $E > 2$ keV is 5.3
$\pm$ 1.3\%, somewhat higher than the modulation found in the lower
energy band ($ E < 2$ keV): 2.7 $\pm 1.2$\%.  Above 2 keV, the hard
power law dominates the spectrum (see section 4.2) while below 1 keV,
the power law and blackbody components have about equal contributions
(see Fig. 4), so this marginally significant change of the pulse
fraction could be obtained if the blackbody component is totally
unpulsed.

\section{Spectral Analysis}

The preparation of the spectra for analysis was done in several
steps. In order to perform the basic spectral fits, we first
subtracted the background from each spectrum and fitted the HEG and
MEG spectra simultaneously using the XSPEC v11.0 package (Arnaud
1996). We also co-added all background-subtracted spectra and exposure
maps and divided the spectra by these exposure maps in order to screen
the fit against local deviations and adjust the fit.  The exposure map
consists of the aspect-corrected effective area of the instrument over
the total exposure for each dispersion arm.  The source spectra were
extracted within a cross-dispersion width of 2$\times 10^{-3}$ degrees
encompassing 98$\%$ of the dispersed flux on the CCD.  We then
selected the first-order spectra from the HEG and MEG. The background
spectra were computed by extracting spectra above and below the
dispersed flux, again with the same total width in
cross-dispersion. The background usually appeared extremely weak,
although we observed some enhanced flux in the MEG -1st order range
between 16 and 20~\AA, at a level about a factor of 2.5 above the
average. Figure 1 shows the background-subtracted count spectrum after
co-adding the MEG and HEG spectra. The lower curve shows the total
contributing background above 10~\AA.

\subsection{Photoelectric Absorption}

Photoelectric absorption by neutral metals along our line of sight
through the interstellar medium attenuates the low-energy X-ray continuum.
With HETGS, we are able to resolve the individual photoelectric absorption
edges of Ne-$K$ (14.3 \AA), Fe-$L_{\rm III}$ (17.5 \AA), and O-$K$ (23.3 \AA),
allowing separate determinations of the column density for each of
these elements (see Table 1).  Figure 2 shows the Ne $K$ and Fe
$L$-edges.  We note that the uncertainties in these edges are still
large, since the derived absolute edge depths are of the order of the 
statistical uncertainties of the data bins, and their determination 
therefore also depends on the proper placement of the continuum (see below).

%%%%%%%%%%%
\centerline{\epsfig{file=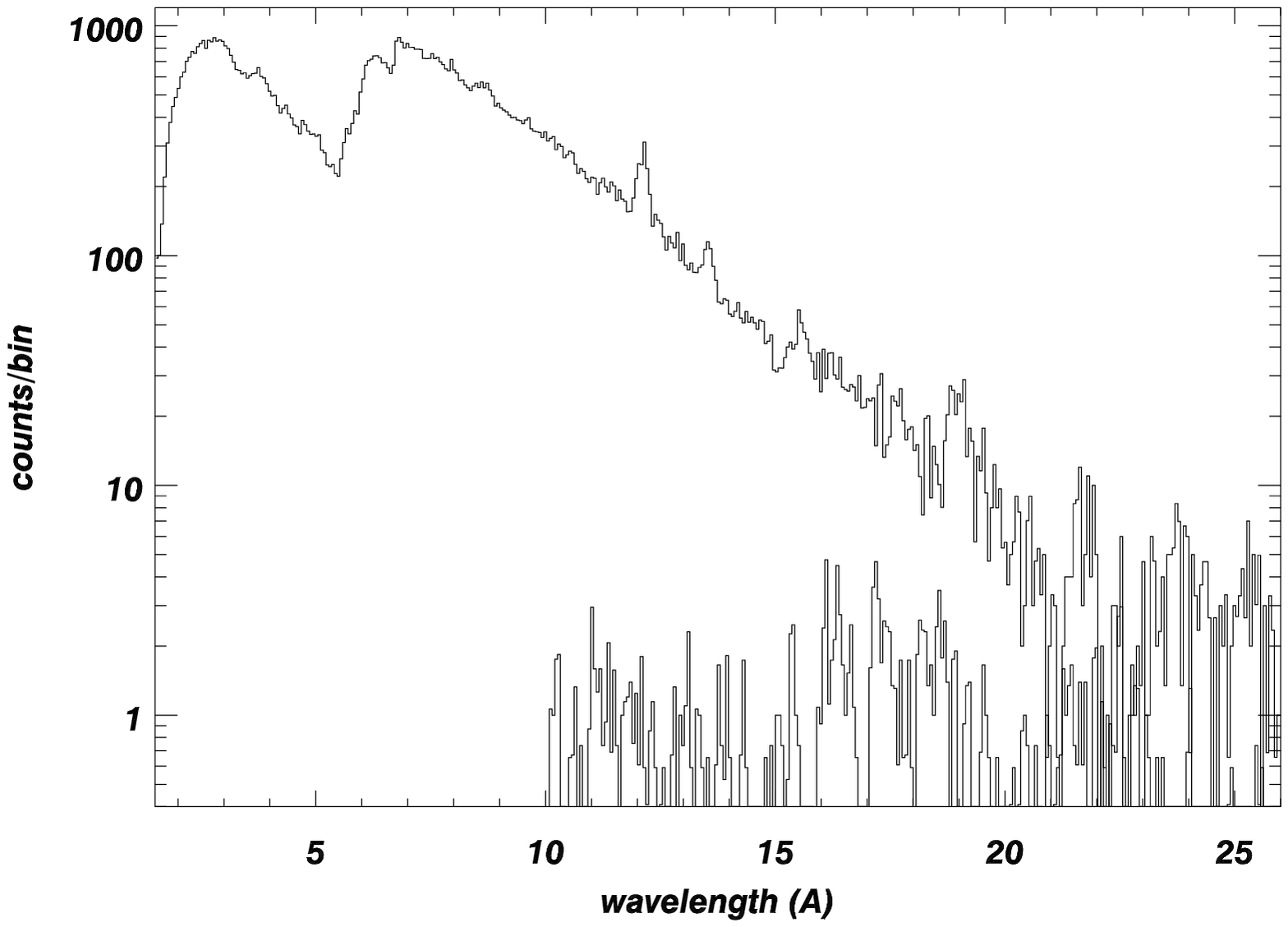,width=0.9\linewidth}}
\figcaption{The {\em Chandra}/HETGS count spectrum of 4U 1626$-$67.
The upper curve is the total background-subtracted spectrum, and the
lower curve is the background above 10~\AA.}
\vspace*{0.1in}
%%%%%%%%%%%
For Ne and O, we used the atomic cross sections of Verner et
al. (1993) but for Fe-$L$ we instead used more recent laboratory
atomic constants (Kortright \& Kim 2000; see also discussion in Schulz
et al. 2001).  Also given in Table 1 is the column density of neutral
hydrogen ($N_{\rm H}$) implied by each element's absorption edge,
assuming the abundances of Morrison \& McCammon (1983) for the material
along the line of sight.  It is interesting to note that for Ne and O,
these implied values of $N_{\rm H}$ are higher than the value of
$N_{\rm H}= (6.2\pm 0.7)\times 10^{20}$ cm$^{-2}$ measured directly
from Ly$\alpha$ measurements in the ultraviolet (Wang \& Chakrabarty
2001).  The abundance ratios also differ substantially from solar values.

Although the threshold for the C-$K$ edge (43.7 \AA) lies beyond our
passband, the short wavelength ``tail'' of this edge does affect the
continuum we observe.  One way to obtain a better representation of the
continuum in the wavelength range above 18 ~\AA\ is to assume a C/H
number ratio (5--18)$\times$ solar.  For our continuum representations
in the figures below, we have conservatively assumed a carbon
overabundance of (C/H)/(C/H)$_\odot$=5.  Given that $N_{\rm H}\lesssim 3\times
10^{21}$ cm$^{-2}$, we expect no sensitivity to absorption edges of Mg
and higher-$Z$ elements.  However, it is interesting to note the
presence of a marginal edge-like feature near the position of the Mg
$K$-edge at 9.5 \AA.  In order to explain this feature by Mg
absorption, we would require an order of magnitude overabundance
of Mg relative to the solar value. 

%%%%%%%%%%%
\vspace*{0.15in}
\centerline{\epsfig{file=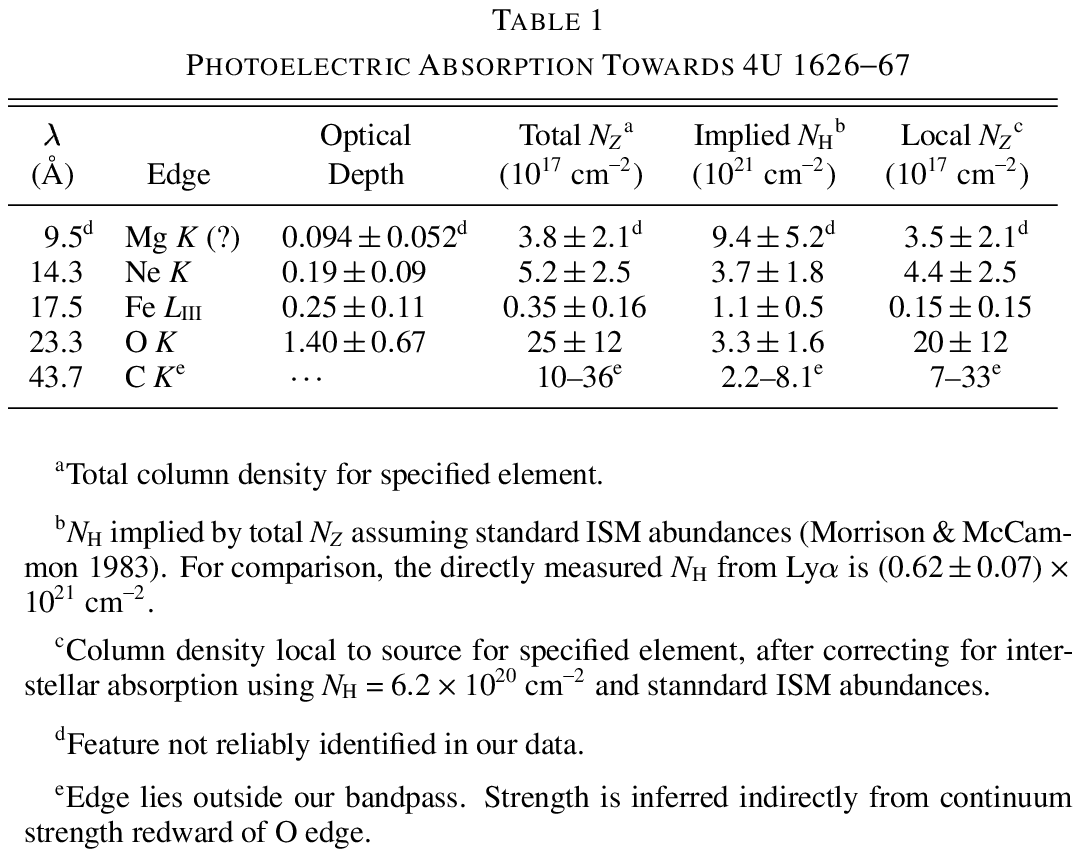,width=0.95\linewidth}}
%%%%%%%%%%%

\begin{figure*}[t]
\centerline{\epsfig{file=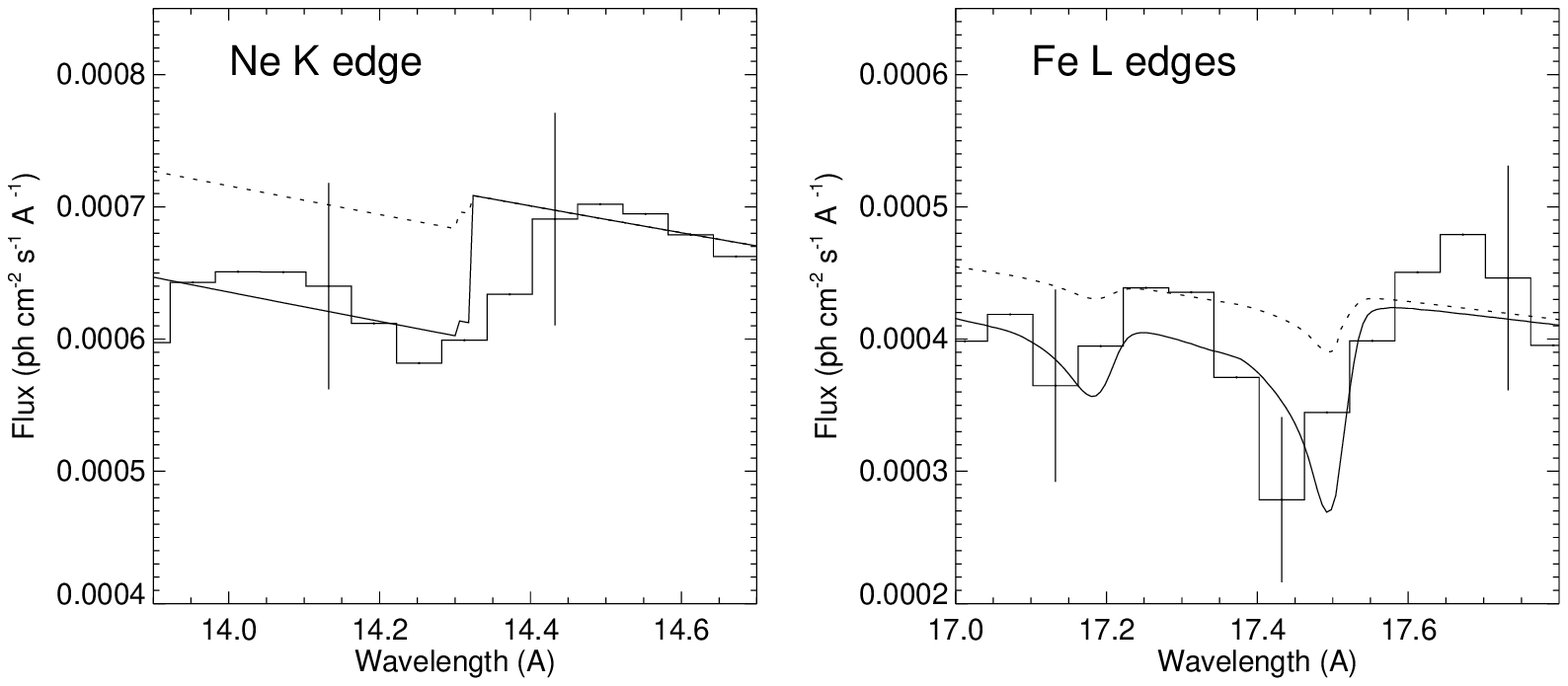}}
\figcaption{The spectrum of 4U 1626$-$67 near the Ne $K$-edge ({\em
left}) and the Fe $L$-edges ({\em right}). The data have been smoothed
for illustrative purposes.  The dotted lines show the expected
contribution of interstellar photoelectric absoprtion based on an
ultraviolet measurement of $N_{\rm H}$.  The solid lines show the
best-fit model for the much stronger absorption edges in the {\em
Chandra} data.} 
\end{figure*}

\begin{figure*}[b]
\centerline{\epsfig{file=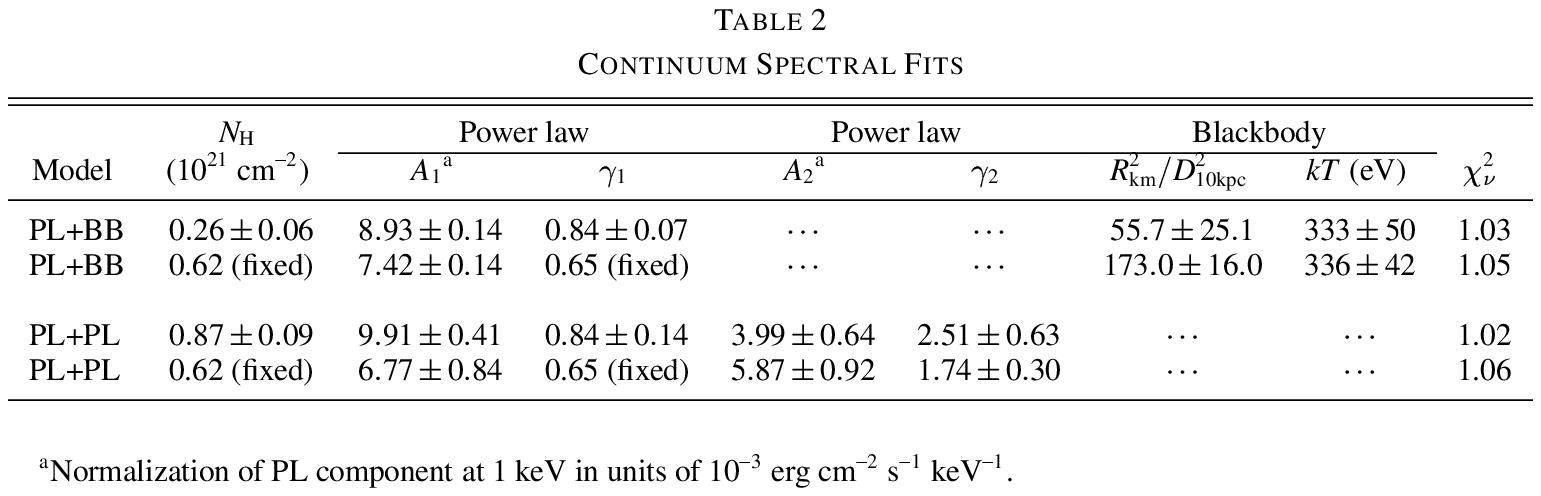}}
\end{figure*}

\subsection{Continuum Model and Fluxes}

To fit the continuum spectrum, we rebinned the data into broad (0.08
\AA) bins, resulting in about 300 wavelength channels.  To avoid
skewing of the fit due to obvious bright lines, we always included
several Gaussian line components in the model fit. This task appeared
to be quite difficult, since we observe quite a number of lines (see
below) in the spectrum, which appear very broad and a substantial
uncertainty remains due to the fact that we cannot identify possible
contributions from many more weak broad lines.  Spectral models with a
single absorbed continuum component did not fit the data.  However,
absorbed two-component fits were successful using either a pair of
power-law components or a sum of power-law and blackbody components.
The best-fit parameters are given in Table 2.  Although the HETGS data
does not clearly prefer one model over the other, the PL+BB model is
consistent with previous low-resolution broadband measurements
(Angelini et al. 1995, Owens et al. 1997), and we will adopt this
model with the photon index fixed at 0.65 for the remainder of this
paper. The spectrum versus wavelength is shown in Figure 3.  A plot of
the photon spectrum versus energy is shown in Figure 4, with the
individual continuum components clearly indicated.

The total measured X-ray flux is 2.09$\times 10^{-10}$ \ergcm (0.5--10
keV).  This corresponds to an (unabsorbed) luminosity of 2.6$\times
10^{34}$ \ergsec $d^2_{kpc}$, which is similar to that reported by
Owens et al. (1997).  This would seem to break the trend of steady
dimming over the entire source history (Chakrabarty et al. 1997).  
The blackbody temperature is also consistent with that measured by
Owens et al. (1997). 

\subsection{Emission Lines}

In order to study discrete line features, we subtracted the continuum
model from the measured photon spectrum. This residual spectrum is
shown at the bottom of Figure~3. It shows a variety of emission and
absorption features.  We have chosen not to analyze the wavelength
range below 8~\AA, since the possible features are weak and are not
easily disentangled from calibration uncertainties. The residual spectrum
contains several strong emission line features at higher wavelength,
which we identify exclusively from hydrogenic and He-like Ne and O ions.
Fitting single Gaussians to the lines yielded line widths ranging from
0.25~\AA\ at \ion{Ne}{10} to 0.48~\AA\ at \ion{O}{7}, corresponding to
Doppler velocities of 5340 and 7460 km s$^{-1}$, respectively.  The
centroid of the \ion{Ne}{9} lines is closest to the expected position of the
intercombination line, while the \ion{O}{7} line lies between the
expected positions for the intercombination and resonance lines.  

Single Gaussian line profiles did not provide an acceptable fit to the
observed line shapes.  Figure 5 shows the profiles of the neon and
oxygen lines, using spectral binnings of 0.015~\AA\ and
0.06~\AA.  The H-like lines clearly show a double line profile.  We 
therefore fit all observed line profiles with Gaussian line doublets.
The resulting line parameters are given in Table 3. For the He-like
triplet lines, we assumed that the individual components of each
triplet were subject to the same velocity shift, and we also enforced
an upper limit on the line flux using the 3 $\sigma$ error level of the
continuum.  We also can further set a lower limit for these fluxes from
calculations for photoionized plasmas by Porquet and Dubau (2000);
specifically, the lines of both \ion{Ne}{9} and \ion{O}{7} should obey
$G\lesssim 4.9$ (see below).  

For each line pair, we denote the higher and lower wavelength components
as ``red'' and ``blue'', respectively, under the assumption that the
wavelength offsets are caused by Doppler redshifts and blueshifts.
Table~3 includes the measured wavelengths and shifts for each of the
observed lines.  We observe blueshifts of 1500--2600 km s$^{-1}$ and
redshifts of 800--1900 km s$^{-1}$.  The FWHM line widths for both the
red and the blue components are roughly 2500 km s$^{-1}$ for all
lines.  The equivalent widths for the integrated red+blue line pairs
are 22.5 eV for \ion{Ne}{10}, 17.8 eV for \ion{Ne}{9}, 24.6 eV for
\ion{O}{8}, and 44.4 eV for \ion{O}{7}.

\subsection{Plasma Diagnostics}

The relative strengths and shapes of the line components in a He-like
triplet are a sensitive diagnostic of plasma temperature and density
(Gabriel \& Jordan 1973; Porquet \& Dubau 2000).  We denote the
resonance, intercombination, and forbidden components as $r$, $i$, and
$f$, respectively.  We also denote hydrogenic Lyman $\alpha$ lines as
$h$.  Plasma temperature may be estimated from the ratio $G=(f+i)/r$.
At high ($\gtrsim 10^6$ K) temperatures, there is an additional
dependence on the ratio of hydrogenic and He-like ions, $H=h/r$.
Plasma density may be inferred from the ratio $R=f/i$.  We compute
these ratios for each of our red and blue line components separately,
and then compare them with the calculations of Porquet \& Dubau (2000)
to infer plasma temperature and density.  The results are summarized
in Table 4.

We infer temperatures of (1.6--1.8)$\times 10^6$ K from the
\ion{Ne}{9} lines and (0.9--1.0)$\times 10^6$ K from the \ion{O}{7}
lines.  For our density estimates, the $R$-value calculations of Porquet \&
Dubau were in fact made for temperatures close to these, 
although our $R$-values are low enough to be fairly insensitive to
temperature in any case.  The density of the \ion{Ne}{9} regions is
about an order of magnitude higher than in the \ion{O}{7} regions.  

Our observation of $r<i$ for the He-like lines would normally be
interpreted as indicative of a purely photoionized plasma, in contrast
to a hybrid plasma which experienced collisional ionization as well;
however, this seems surprising for the high densities we measure.  The
calculations of Porquet and Dubau (2000) do not include scattering
effects, which may be important in high-density regimes.  From the
fact that $G>1$ is well established for all of our line components, we
can deduce that the emission arises from a recombining plasma
(A. Pradhan 2001, personal communication; Liedahl et al. 2001). Given
that we observe moderately high densities in a hot recombining plasma,
it seems likely that collisional ionization is indeed still
important. The weak resonance line in the He-like triplets may
therefore be an opacity effect due to enhanced line scattering.
\begin{figure*}
\centerline{\epsfig{file=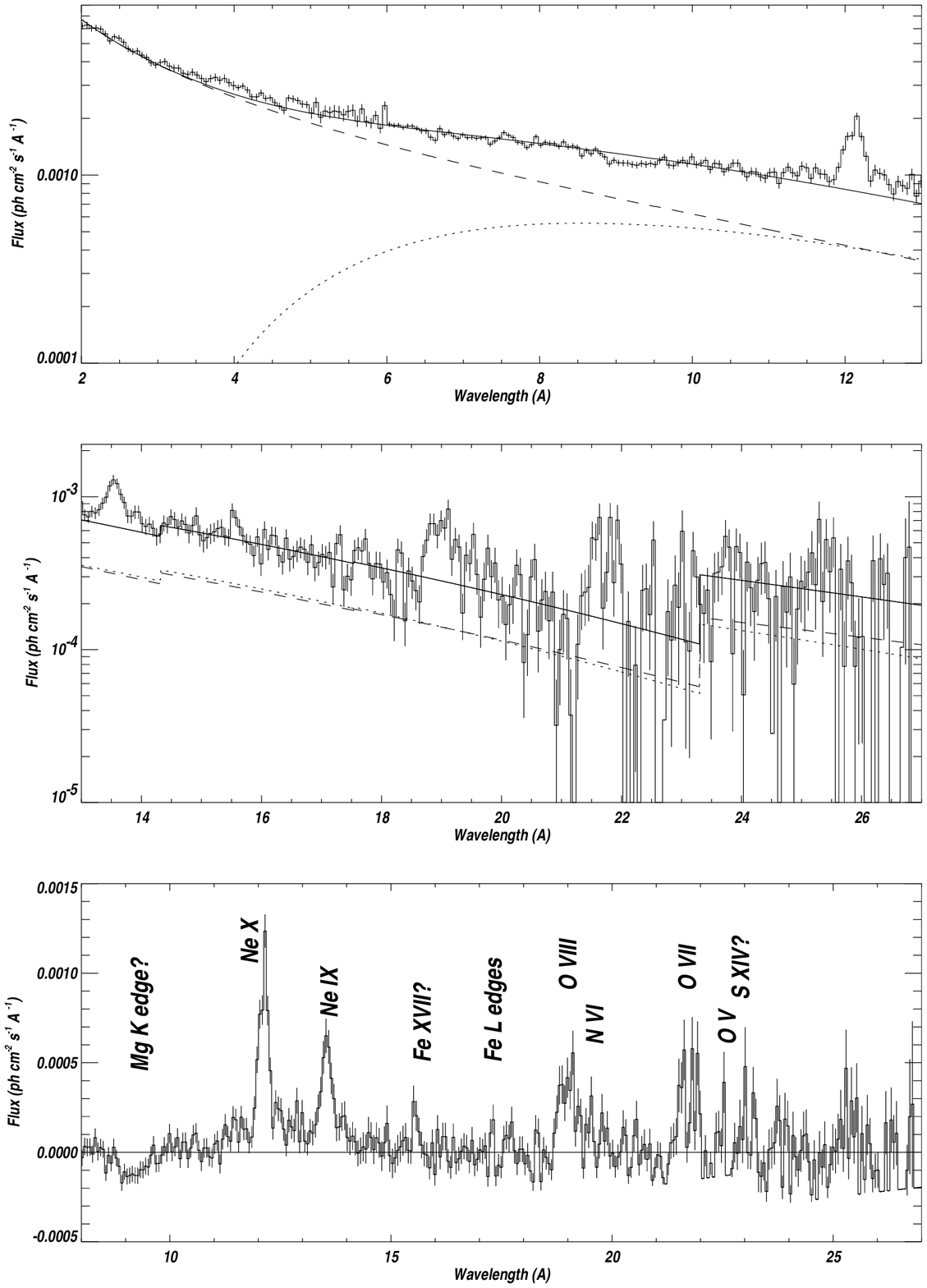}}
\figcaption{The HETGS photon spectrum of 4U 1626$-$67 as a function of
wavelength, along with the best-fit power law (dashed line) and
blackbody (dotted line) components.  The residual spectrum in the
bottom panel identifies various broad emission lines as well as
ionization edges.}
\end{figure*}

\begin{figure*}[b]
\centerline{\epsfig{file=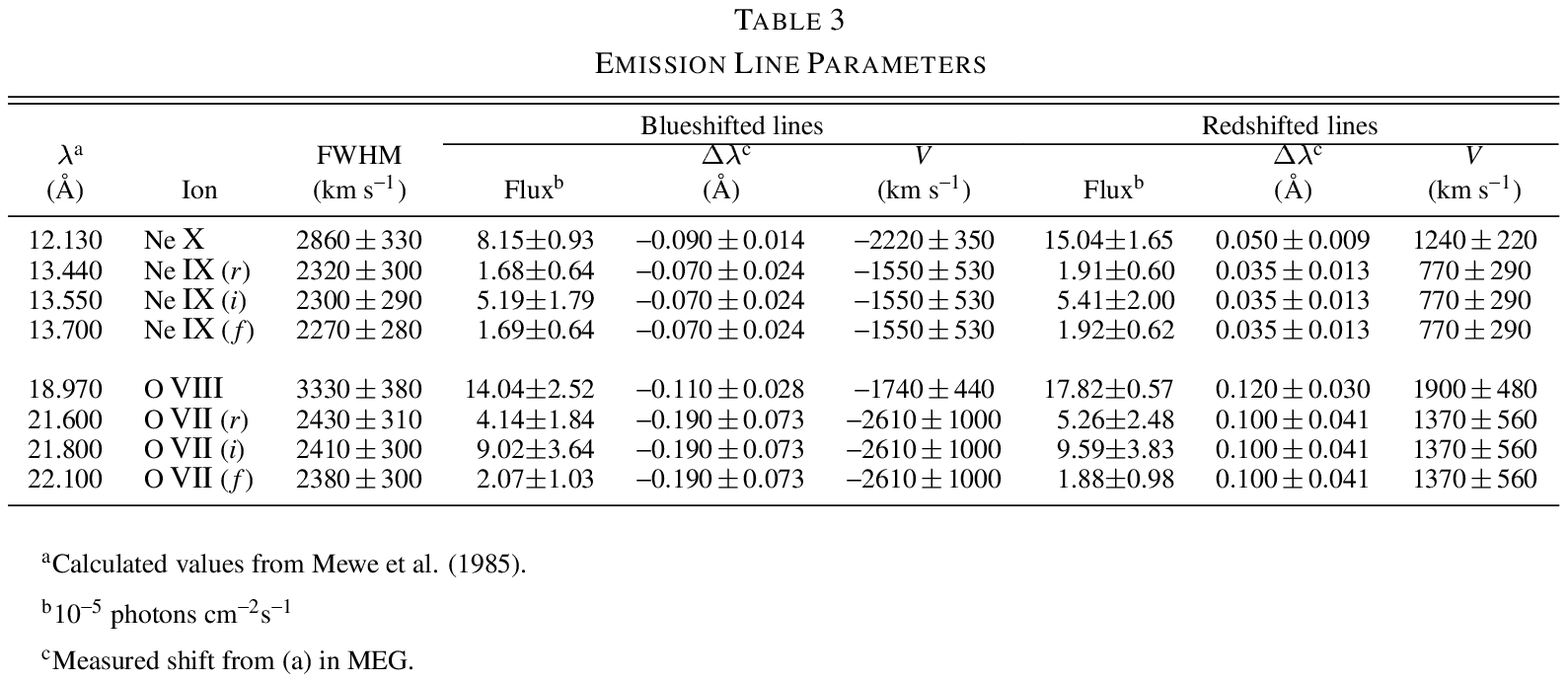}}
\end{figure*}

%%%%%%%%%
\centerline{\epsfig{file=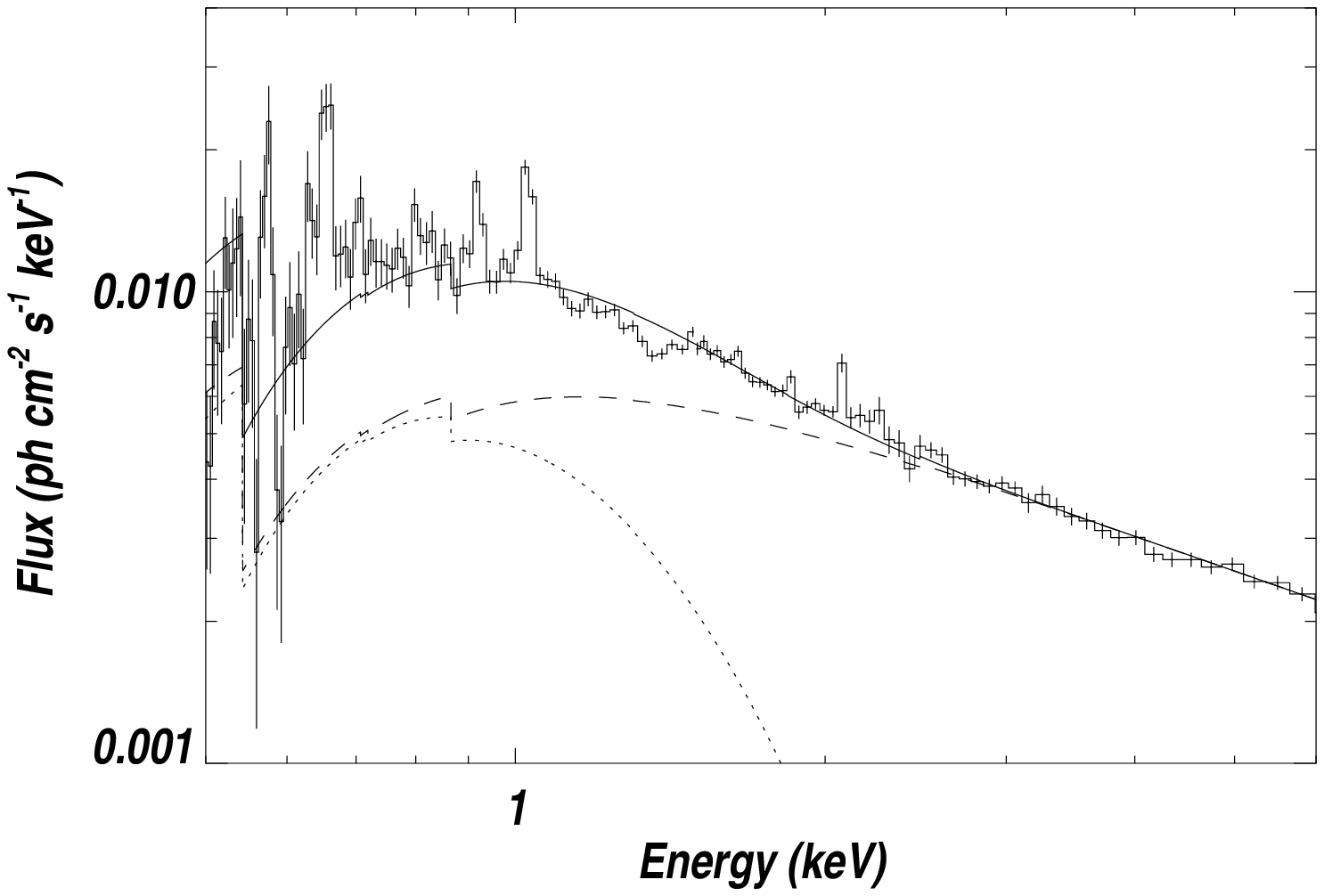,width=0.95\linewidth}}
\figcaption{The broad-band photon spectrum of 4U 1626$-$67 as a
function of energy, showing the relative contributions of the
blackbody (dotted line) and power law (dashed line) components. The
power law component clearly dominates at high energy.} 
\vspace*{0.15in}

\noindent
This introduces some additional uncertainty to our measured $G$-value and
the inferred temperature.

\section{Discussion}

\subsection{Origin of the Double-Peaked Emission Lines}

We have resolved the previously detected Ne/O emission line complex
into Doppler pairs of broadened lines from hydrogenic and He-like Ne
and O.  For most of the line pairs, the red and blue components have
comparable strength (the one exception being the \ion{Ne}{10} line).
X-ray line complexes with well-separated red/blue line pairs have been
previously detected from relativistic jets in the so-called Galactic
microquasars SS~433 (Kotani et al. 1996; Marshall, Canizares, \&
Schulz 2001) and 1E 1740.1-2947 (Cui et al. 2000), but the Doppler
shifts in those cases are relativistic, quite unlike what we observe
in 4U 1626$-$67.  Instead, the velocity width of the lines and their
projected Doppler velocity shifts are of the same order of magnitude
as the Keplerian velocities in the accretion disk (970--5400
km~s$^{-1}$). Thus, the most natural explanation for these lines is that
they arise in or near the Keplerian disk flow around the neutron
star. Similar double line profiles are frequently observed in
optical/UV emission lines from accretion disk flows in cataclysmic
variables (see, e.g., Frank, King, \& Raine 1992), but our observation
would be the first X-ray detection of such features.  (Single-peaked
lines with comparable broadening and He-like line ratios have been 
observed in the edge-on LMXB EXO 0748$-$676; Cottam et al. 2001).  We
note that if this Keplerian explanation is correct, then the companion
must be a 0.02~$M_\odot$ white dwarf.  The other alternative (a
hydrogen-depleted, partially degenerate 0.08 $M_\odot$ star) requires
a very small ($\lesssim 8^\circ$) inclination.  This gives a maximum
projected velocity at the inner disk edge of only $<750$ km~s$^{-1}$,
inconsistent with our observed Doppler shifts.

\begin{figure*}[t]
\centerline{\epsfxsize=8.5cm\epsfbox{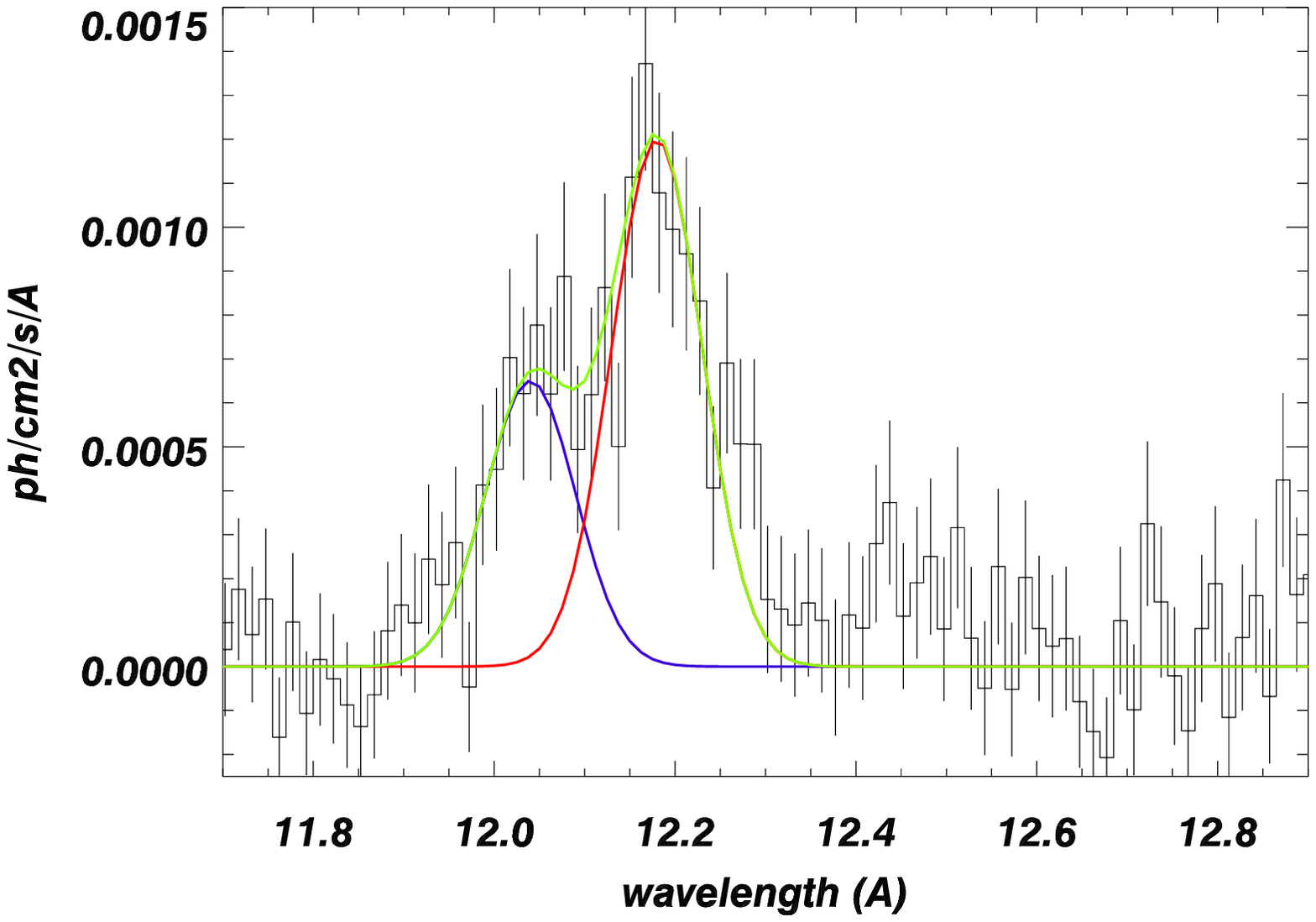}\epsfxsize=8.5cm\epsfbox{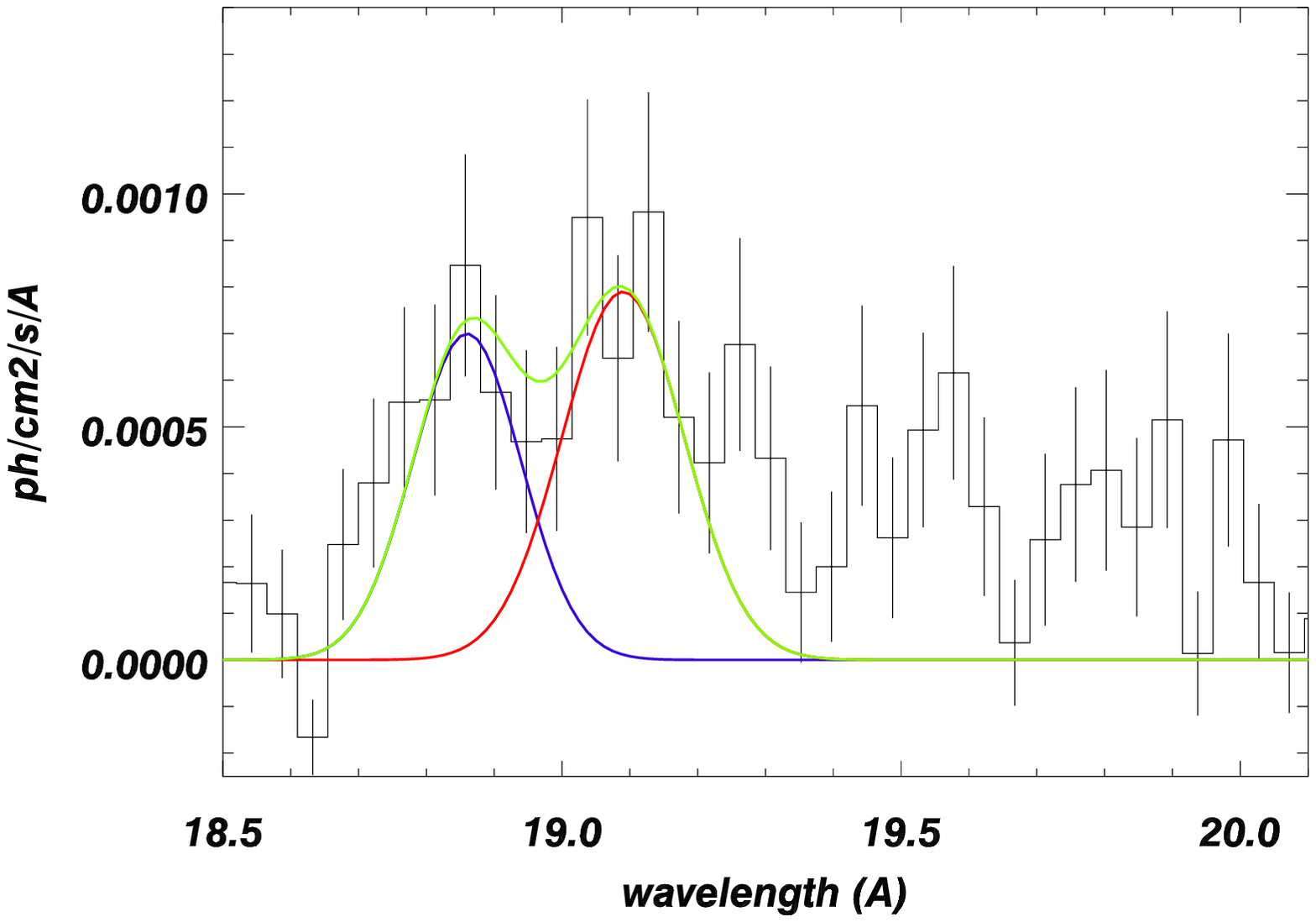}}
\centerline{\epsfxsize=8.5cm\epsfbox{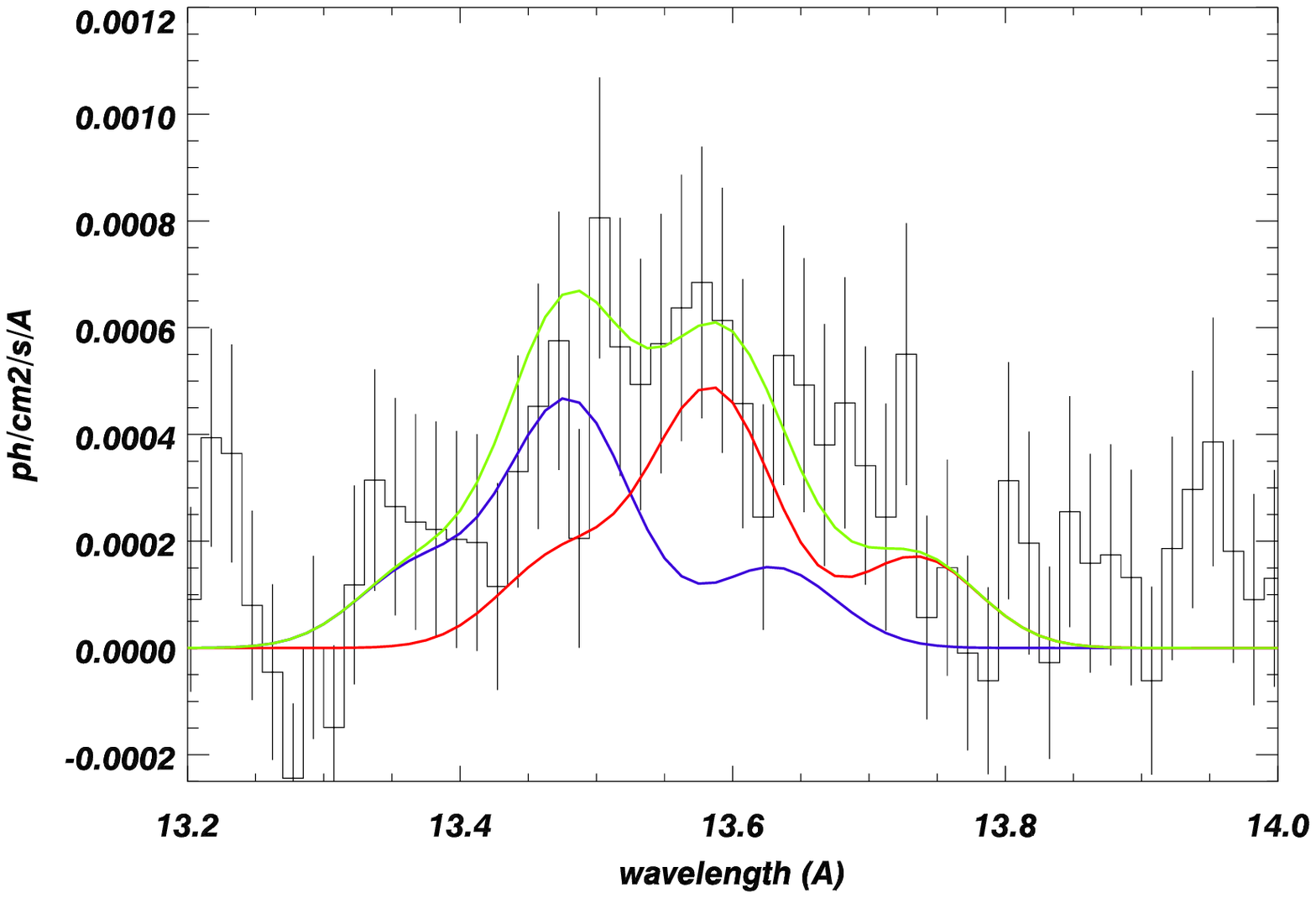}\epsfxsize=8.5cm\epsfbox{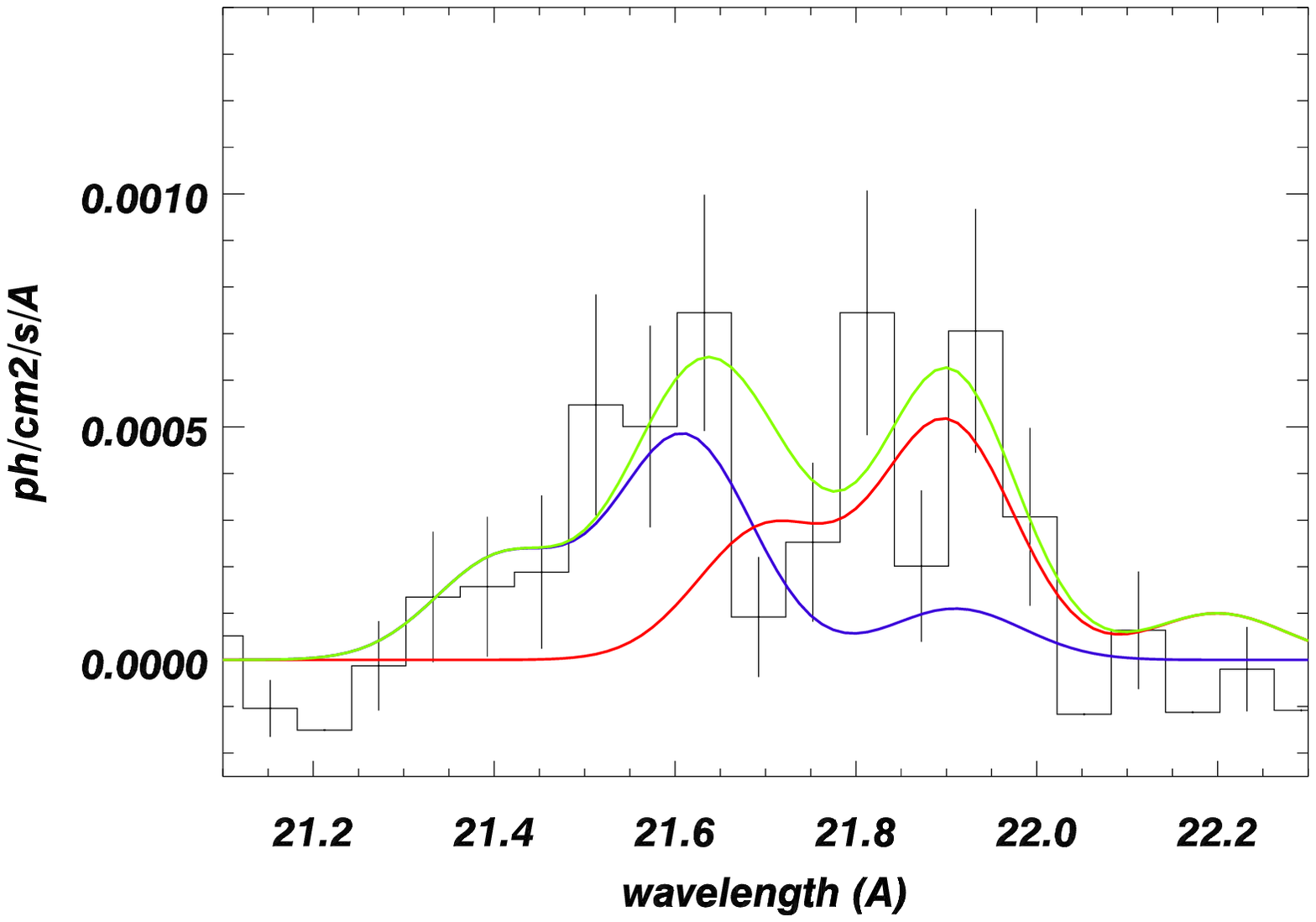}}
\figcaption{Double-peaked, broad emission lines from 4U 1626$-$67.
The upper panels show the hydrogenic Ne~X ({\em upper left})
and O~VIII ({\em upper right}) lines, and the lower panels show
the He-like triplets of Ne~IX ({\em lower left}) and O~VII
({\em lower right}).  The blue curves denote the model for the
blue-shifted components, the red curves the model for the red-shifted
components, and the green curves the model for the overall line shape.
In addition to the Doppler shifts, the individual line components are
all broadened ($\approx 2500$~km~s$^{-1}$ FWHM).}
\end{figure*}

Alternatively, if the companion really is a more massive star, then we
are viewing the disk nearly face-on and the Doppler line pairs are
more difficult to understand.  One possible explanation is that they
arise in a bipolar outflow.  Several authors have proposed that a
magnetohydrodynamic wind might be driven from the magnetopause (near
the inner disk edge) of an accreting neutron star during spin down
(Anzer \& B\"orner 1980; Arons et al. 1984; Lovelace, Romanova, \&
Bisnovatyi-Kogan 1995).  Such an outflow would presumably have roughly
the Keplerian velocity at the inner disk edge but would be oriented at
a substantial angle to the disk plane.  An advantage of this picture
is that there is already indirect evidence that some outflow from the
disk must be present, since the X-ray flux from the source (which
traces $\dot M$ onto the neutron star) has been steadily dropping for
years while the optical flux (which traces $\dot M$ through the disk)
has remained unchanged.  We can estimate the wind mass loss rate as
$-\dot M_{\rm wind}\sim A\,n_e\,m_p\,r^2\,V\simeq 10^{-10} M_\odot$
yr$^{-1}$, where we have used the inner disk radius and Keplerian
velocity and have taken $A=16$ for oxygen.  This is of the same order
as the observed mass accretion rate onto the neutron star.  However, a
serious drawback of this explanation is that it fails to account for the
comparable strengths of the red and blue line components, since we
would expect the redshifted lines to be at least partially blocked
from our line of sight by the accretion disk itself.  Consequently, we
do not regard this explanation as likely.

The He-like triplets of O and Ne are dominated by the intercombination
component, indicating that the lines arise in a relatively dense,
photoionized plasma (see, e.g., Liedahl et al. 2001).   The inferred plasma
temperatures (see Table 4) are considerably higher than the expected
surface temperature range of the optically thick accretion disk.
However, detailed calculations of the structure of X-ray heated
accretion disks indicate the existence of three sharply distinct
layers near the surface: an X-ray heated, fully-ionized outer layer at the
Compton temperature ($\sim10^7$--$10^8$ K); a thin, highly-ionized layer
just below that, with $T\sim10^6$K; and the usual, optically thick
disk surface below that, with $T\sim 10^4$--$10^5$K (Nayakshin,
Kazanas, \& Kallman 2000; Li, Gu, \& Kahn 2001).  Our measured
temperatures and ionization states are thus consistent with those
expected in the narrow, highly ionized layer just below the X-ray
heated skin.  For the observed O and Ne ion species to survive, the
ionization parameter $\xi=L/nR^2$ must be below $10^3$ erg cm$^{-2}$ s$^{-1}$
(Kallman \& McCray 1982), where $n$ is the ion number density.   This
gives a rough lower limit of $\gtrsim 1\times 10^{10}\,d_{\rm kpc}$ cm
for the radius of the line emitting region.  This is consistent with
an origin in the outer accretion disk, where X-ray heating effects
dominate.    We note that, for the inferred temperatures, plasma
simulations with the XSTAR v2 code (Kallman \& McCray 1982; Kallman \&
Bautista 2000) predict line emission from Si and S which we do not
observe.  This again points to an overabundance of Ne and O relative to
solar values, as previously found by Angelini et al. (1995).

\subsection{Absorption Edges and the Nature of the Mass Donor}

Several of the observed photoelectric absorption edges are
considerably stronger than expected for interstellar material, given
the modest neutral hydrogen absorption determined from Ly$\alpha$
measurements.  We note that the most clearly detected strong edges are
those of Ne and O, the same elements whose emission lines we discussed
above.  We thus conclude that the strong edges are due to absorption
in cool, metal-rich material local to the source.  (We may neglect
local contributions to the hydrogen column density, as we know the
donor in an ultracompact binary must be hydrogen-depleted; see Nelson
et al. 1986.)  Angelini et al. (1995) raised the possibility that 4U
1626$-$67 is embedded in the supernova remnant from the birth of the
pulsar, but as they point out, the absence of strong lines from Si, S,
and Fe is inconsistent with this idea.  Instead, we presume that this
cool material originated in the accretion disk or the mass donor, so
that its composition is then a clue to the nature of the companion.
In the last column of Table~1, we calculate the column density of
material local to the source, assuming that the interstellar medium
contributes $N_{\rm H}=4.5\times 10^{20}$ with standard elemental
abundances.  The local column density of the element with atomic
number $Z$ is then $N_Z^{\rm loc} = N_Z^{\rm tot} - N_{\rm H}A_Z$,
where $N_Z^{\rm tot}$ is the {\em total} measured column density of
that element and $A_Z$ is the number abundance of the element relative
to hydrogen in the interstellar medium.

\begin{figure*}[t]
\centerline{\epsfig{file=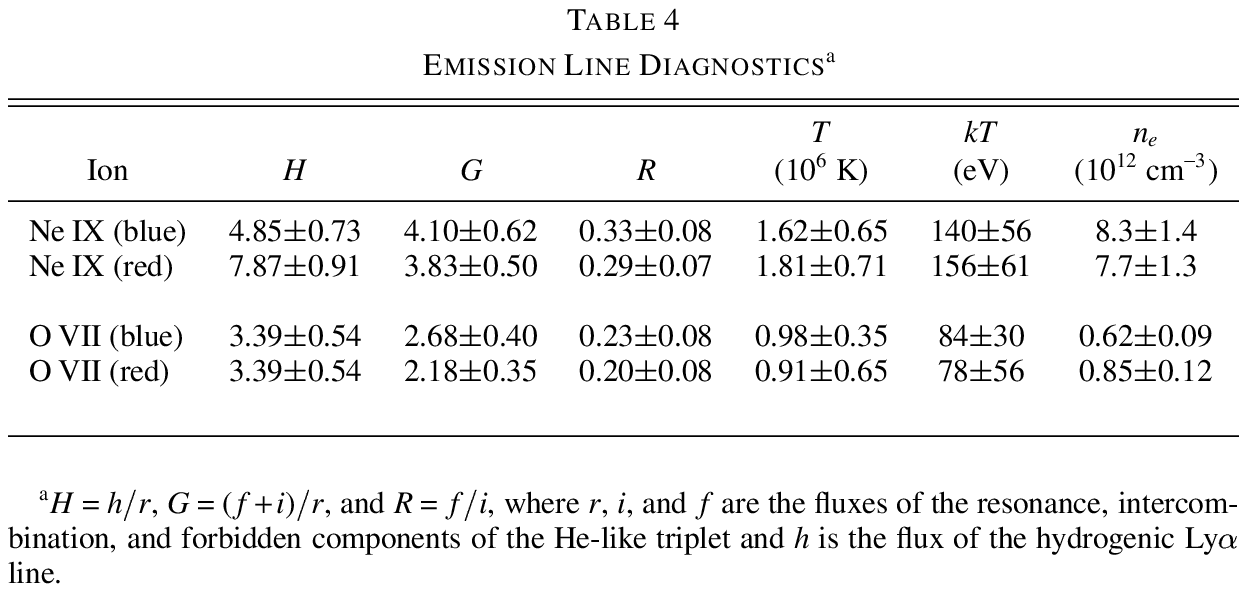}}
\end{figure*}

We may compare our resulting abundance ratios with those expected for
the possible companion types.  The Ne/O abundance ratio of $0.22\pm
0.15$ is much larger than expected for either a partially degenerate
hydrogen-depleted star or a He white dwarf.  On the other hand,
$^{22}$Ne is the next most abundant element in a C-O white dwarf,
which has typical initial mass fractions of $X_{\rm C}=X_{\rm O}=0.49$
and $X_{\rm Ne}=0.02$ (Segretain et al. 1994). This still gives a Ne/O
ratio of only a few percent.  However, a white dwarf that has cooled
sufficiently will undergo crystallization (see, e.g., Shapiro \&
Teukolsky 1983), leading to chemical differentiation (Isern et
al. 1991).  In a C-O dwarf, this will cause all the Ne nuclei to settle into
the inner few percent of the star and will also cause an O enhancement
(Isern et al. 1991); detailed calculations predict $X_{\rm C}\approx
0.19$, $X_{\rm O}\approx 0.65$, and $X_{\rm Ne}\approx 0.16$ (Segretain
et al. 1994), consistent with our observed ratios.  Note that since a
mass of 0.02 $M_\odot$ is required for a white dwarf secondary in this
system, it is precisely the inner few percent of the star that is
relevant, since the outer layers of the white dwarf would have been
stripped away much earlier in the binary's accretion history.  The C/O
abundance ratio is poorly constrained by our observations, but the
range of allowed values is consistent with the expected ratio in the
core of C-O dwarf that has crystallized.

If the feature near 9.5~\AA\ is in fact the Mg $K$-edge, then there is
also a substantial overabundance of Mg relative to the solar value.  The
resulting Mg/O abundance ratio ($0.17\pm 0.13$) would be quite high
for a C-O dwarf, and instead suggests the possibility that the
companion is an O-Ne-Mg dwarf.  Although these stars have a typical
initial composition of $X_{\rm O}=0.72$, $X_{\rm Ne}=0.25$, and
$X_{\rm Mg}=0.03$ (see Gutierrez et al. 1996 and references therein),
crystallization during cooling would likely lead to a substantially
enhanced central concentration of Mg, by analogy to what is calculated
for C-O dwarfs (Ogata et al. 1993).   Some models suggest that the
central region might also contain some unburned $^{12}$C from when the
O-Ne-Mg dwarf is formed (Dominguez, Tornambe, \& Isern 1994), which might
account for the C overabundance suggested in our analysis. 

Taken together, the double-peaked emission lines and strong absorption
edges of Ne and O in our HETGS spectrum thus point to the companion in
4U 1626$-67$ being the chemically fractionated core of a C-O or
O-Ne-Mg crystallized white dwarf, the first example of such a donor
among the low-mass X-ray binaries.  In either case, the progenitors of
both components of the binary must have been massive stars in an
initially wide orbit in order to avoid a common envelope phase prior
to the formation of a C-O or O-Ne-Mg degenerate core (see Levine et
al. 1988 and references therein).  The extremely low mass of the donor
suggests an extended period of mass transfer.  However, it is
generally believed that prolonged mass transfer onto neutron stars
leads to millisecond spin periods and weak ($\sim 10^8$ G) magnetic
field strengths (e.g., Bhattacharya \& van den Heuvel 1991).  By
contrast, 4U 1626$-$67 has a slow (7.66 s) spin period and a strong
($\sim 10^{12}$ G) magnetic field.  One possible explanation is that
the neutron star is relatively young and that it formed from
accretion-induced collapse of a massive white dwarf (Taam \& van den
Heuvel 1986; Levine et al. 1988; Verbunt et al. 1990).  If that is the
case, then the strong absorption edges we observe may actually arise
from material expelled from this progenitor at the time of its
collapse into a neutron star, rather than from the mass donor or the
accretion disk.

One further puzzle is to reconcile the observationally inferred mass
transfer rate of $\dot M\approx 2\times 10^{-10} M_\odot$~yr$^{-1}$
(Chakrabarty et al. 1997; Chakrabarty 1998; Wang \& Chakrabarty 2001)
with theoretical expectations.  Mass transfer in a 42-min binary
is driven by angular momentum losses via gravitational radiation
(Verbunt \& van den Heuvel 1995 and references therein).  For a 0.02
$M_\odot$ donor, we expect only $\dot M\approx 3\times 10^{-11}
M_\odot$~yr$^{-1}$.  It is possible that the currently observed rate does
not reflect the long-term average rate.  Indeed, we note that for
$\dot M\sim 10^{-11} M_\odot$ yr$^{-1}$, a C-O accretion disk may be
subject to dwarf-nova--like thermal ionization instabilities (Menou,
Perna, \& Hernquist 2001).   This may be true for O-Ne-Mg disks as
well, although we are not aware of any explicit calculations.   While
4U 1626$-$67 is generally considered a persistent source, its flux has
been decaying steadily for over a decade.  It is possible that it is
actually a very long-lived X-ray transient, and that its average $\dot
M$ (calculated over many decades) is consistent with the theoretically
predicted value.

\acknowledgments It is a pleasure to thank Lars Bildsten for useful
discussions, particularly on the subject of white dwarf composition
and crystallization.  The authors also thank the MIT HETG instrument
team and the {\em Chandra} X-Ray Center for their support.  This
research was supported by contracts SAO SV1-61010 and NAS8-38249 as
well as NASA grant NAG5-9184.


\begin{references}
\reference{Ange1995} Angelini, L., White, N.~E., Nagase, F., Kallman, T.~R.,
Yoshida, A., Takeshima, T., Becker, C., \& Paerels, F. 1995, \apj, 449
L41 

\reference{ab80} Anzer, U. \& B\"orner, G. 1980, \aap, 83, 133

\reference{arnaud96} Arnaud, K.~A. 1996, in Astronomical Data Analysis
Software and Systems V, ed. G. Jacoby \& J. Barnes (San Francisco: ASP
Conf. Ser. 101), 17

\reference{arons84} Arons, J., Burnard, D., Klein, R.~I., McKee,
C.~F., Pudritz, R.~E., \& Lea, S.~M. 1984, in High Energy Transients
in Astrophysics, ed. S.~E. Woosley (New York: AIP Press), 215


\reference{bha91} Bhattacharya, D. \& van den Heuvel, E.~P.~J. 1991,
Phys. Rep., 203, 1

\reference{chak1998} Chakrabarty D., 1998, \apj, 492, 342 

\reference{chak1997} Chakrabarty D. et al. 1997, \apj,  474, 414 

\reference{cott2001} Cottam, J., Kahn, S.~M., Brinkman, A.~C., den
Herder, J.~W., \& Erd, C., 2001, \aap, 365, L277

\reference{cui2001}  Cui, W., Schulz, N.~S., Baganoff, F.~K., Bautz,
M.~W., Doty, J.~P., Garmire, G.~P., Mirabel, I.~F., Ricker, G.~R., 
Rod\'riguez, L.F., \& Taylor, S.~C. 2001, \apj, 548, 394

\reference{dti94} Dominguez, I., Tornambe, A., \& Isern, J. 1994,
\apj, 419, 268

\reference{elsn1983} Elsner, R.~F., Darbro, W., Leahy, D., Weisskopf, M.~C.,
Sutherland, P.~G., Kahn, S.~M., \& Grindlay, J.~E. 1983, \apj, 266, 769 

\reference{fkr92} Frank, J., King, A., \& Raine, D. 1992, Accretion
Power in Astrophysics, 2nd ed. (Cambridge: Cambridge U. Press)

\reference{gj73} Gabriel, A.~H. \& Jordan, C. 1973, \apj, 186, 327

\reference{guti96} Gutierrez, J., Garcia-Berro, E., Iben, I., Isern,
J., Labay, J., \& Canal, R. 1996, \apj, 459, 701

\reference{img91} Isern, J., Mochkovitch, R., Garcia-Berro, E., \&
Hernanz, M. 1991, \aap, 241, L29

\reference{kall2000} Kallman, T.~R. \& Bautista, M. 2000, BAAS, 32,
1227 (HEAD abstract 27.03)

\reference{kall1982} Kallman, T.~R. \& McCray, R. 1982, \apjs, 50, 263

\reference{Kii1986}  Kii, T., Hayakawa, S., Nagase, F., Ikegami, T., \&
Kawai, N. 1986, \pasj, 38, 751 

\reference{kk00}  Kortright, J.~B. \& Kim, S.-K. 2000, Phys. Rev. B,
62, 12216

\reference{Kota1996} Kotani, T., Kawai, N., Matsuoka, M., \& Brinkmann,
W. 1996, \pasj, 48, 619 

\reference{levi1988} Levine, A., Ma, C.~P., McClintock, J., Rappaport, S.,
van der Klis, M., \& Verbunt F. 1988, \apj, 327, 732 

\reference{li2001} Li, Y., Gu, M.~F., \& Kahn, S.~M. 2001, \apj, submitted
(astro-ph/0106163)

\reference{lied2001} Liedahl, A.~D., Wojdowski P.~S., Jimenez-Garate, M.~A., \&
Sako, M. 2001, in High-Resolution X-Ray Spectroscopy, ed. G. Ferland
(San Francisco: ASP Conf. Ser.), in press (astro-ph/0105084)

\reference{lrb95} Lovelace, R.~V.~E., Romanova, M.~M., \&
Bisnovatyi-Kogan, G.~S. 1995, MNRAS, 275, 244

\reference{Maur1982} Maurer, G.~S., Johnson, W.~N., Kurfess, J.~D., \&
Strickman, M.~S. 1982, \apj, 254, 271 

\reference{Mars2001} Marshall, H.~L., Canizares, C.~R., \& Schulz, N.~S.
2001, \apj, submitted 

\reference{Mavr1994} Mavromatakis, F. 1994, \aap, 285, 503

\reference{Menou2001} Menou, K., Perna, R., \& Hernquist, L. 2001,
\apj, in press (astro-ph/0102478)

\reference{mew1995} Mewe, R., Gronenschild, E.~H.~B.~M., \& van den Oord,
G.~H.~J. 1985, A\&AS, 62, 197 

\reference{midd1981} Middleditch, J., Mason, K.~O., Nelson, J.~E., \&
White, N.~E. 1981, \apj, 244, 1001 

\reference{morr1983} Morrison, R. \& McCammon, D. 1983, \apj, 270, 119

\reference{naya2000} Nayakshin, S., Kazanas, D., \& Kallman,
T.~R. 2000, \apj, 537, 833

\reference{nels1986} Nelson, L.A., Rappaport, S.~A., \& Joss, P.~C. 1986,
\apj, 304, 231 

\reference{ogata93} Ogata, S., Iyetomi, H., Ichimaru, S., \& van Horn,
H. 1993, Phys. Rev. E, 48, 1344

\reference{orla1998} Orlandini, M., Dal Fiume, D., Frontera, F., del Sordo,
S., Piraino, S., Santangelo, A., Segreto, A.,  Oosterbroek, T., \& Parmar,
A.~N. 1998, \apj, 500, L163 

\reference{owen1997} Owens, A., Oosterbroek, T., \& Parmar A.~N. 1997,
\aap, 324, L9 

\reference{pacz1981} Paczynski, B. \& Sienkiewicz, R. 1981, \apj, 248, 27

\reference{porq2000} Porquet, D. \& Dubau, J. 2000, A\&AS, 143, 495

\reference{prav1979} Pravdo, S.~H. et al. 1979, \apj, 231, 912 

\reference{schu2001} Schulz, N.~S., Cui, W., Canizares, C.~R., Marshall,
H.~L., Lee, J.~C., Miller, J.~M., \& Lewin, W.~H.~G. 2001, \apj, in press
(astro-ph/0109236)

\reference{segre94} Segretain, L., Chabrier, G., Hernanz, M.,
Garcia-Berro, E., Isern, J., \& Mochkovitch, R. 1994, \apj, 434, 641

\reference{st83} Shapiro, S.~L. \& Teukolsky, S.~A. 1983, Black Holes,
White Dwarfs, and Neutron Stars: The Physics of Compact Objects (New
York: Wiley)

\reference{taam86} Taam, R.~E. \& van den Heuvel, E.~P.~J. 1986, \apj,
305, 235

\reference{vaug1997} Vaughan, B.~A. \& Kitamoto, S. 1997, preprint
(astro-ph/9707105)

\reference{vv95} Verbunt, F. \& van den Heuvel, E.~P.~J. 1995, in
X-Ray Binaries, ed. W.~H.~G. Lewin, J. van Paradijs, \& E.~P.~J. van
den Heuvel (Cambridge: Cambridge U. Press), 457

\reference{vwb90} Verbunt, F., Wijers, R.~A.~M.~J., \& Burm,
H.~M.~G. 1990, \aap, 234, 195 

\reference{verner1993} Verner, D.~A., Yakovlev, D.~G., Band, I.~M., \&
Trzhaskovskaya, M.~B. 1993, Atomic Data \& Nucl. Data Tables, 55, 233

\reference{wang2001} Wang, Z. \& Chakrabarty, D. 2001, \apj, submitted
\end{references}
\end{document}